\documentclass[doublecol]{epl2}

\usepackage{graphicx}
\usepackage{epstopdf, epsfig}
\usepackage{amsmath}
\usepackage{amsfonts}
\usepackage{amssymb} 
\usepackage{graphicx}
\usepackage{setspace}
\usepackage{subfigure}
\usepackage{graphicx}  
\usepackage{dcolumn}   
\usepackage{bm}        
\usepackage{amsmath,amssymb}   
\usepackage{color}
\usepackage{enumitem}

\newcommand{\e}{\mathrm{e}}

\definecolor{colortodo}{RGB}{250,0,0}

\title{Capillary wave turbulence experiments in microgravity}
\shorttitle{Capillary wave turbulence experiments in microgravity} 

\author{M. Berhanu \inst{1} \and E. Falcon \inst{1} \and G. Michel \inst{2,3} \and C. Gissinger \inst{2} \and S. Fauve \inst{2}}
\shortauthor{M. Berhanu \etal}

\institute{                  
  \inst{1} Laboratoire Mati\`ere et Syst\`emes Complexes, CNRS UMR 7057 Universit\'e Paris Diderot, 10 rue Alice Domon et L\'eonie Duquet, Paris, France\\
  \inst{2} Laboratoire de Physique de l’Ecole normale sup\'erieure, ENS, Universit\'e PSL, CNRS, Sorbonne Universit\'e, Universit\'e Paris Diderot, Paris, France
    \inst{3} Sorbonne Universit\'e, CNRS, UMR 7190, Institut Jean Le Rond d’Alembert - F-75005 Paris, France
}

\abstract{Using the FLUIDICS (Fluid Dynamics in Space) experiment in the International Space Station, turbulence of capillary waves at the air-water interface is experimentally investigated in weightlessness. Capillary waves are excited in a spherical container partially filled with water and undergoing sinusoidal or random oscillations. The fluctuations of the interface, recorded with two capacitive probes are analyzed by means of the frequency power spectrum of wave elevation. For high enough forcing amplitudes, we report power-law spectra with exponents close to the prediction of weak wave turbulence theory. However, in this experiment the free-surface steepness is not small compared to 1 and thus the investigated regimes correspond to strongly nonlinear wave turbulence.}

\begin{document}

\maketitle

Wave turbulence describes the statistical behavior of a random set of dispersive waves in nonlinear interaction. This phenomenon occurs at very different scales in a great variety of systems~\cite{Nazarenko2011,Newell2011}. The case of surface waves at the interface between a gas and a liquid constitutes an example of prime interest, which has deserved numerous experimental investigations~\cite{Falcon2010,Nazarenko2016}. With specifically the hypotheses of weak nonlinearity and negligible dissipation, wave turbulence theory~\cite{Zakharov1992,Nazarenko2011,Newell2011,Nazarenko2016}
predicts, in stationary state, turbulent self-similar regimes, in which energy (or another conserved quantity) is transferred from an injection scale towards a dissipation scale. These regimes are characterized by power law spectra both in spatial and temporal Fourier spaces, for scales belonging to the inertial range, where both the dissipation and the forcing are negligible. In contrast to hydrodynamic turbulence for which the Kolmogorov spectrum has not been deduced yet from the Navier-Stokes equation, energy spectra can be analytically computed in the framework of weak wave turbulence. Since the hypotheses used in this derivation are drastic, the relevance of weak wave turbulence theory to describe experimental and natural systems remains questionable. In particular, wave turbulence theory assumes a scale-invariant dispersion relation of the form $\omega \propto \vert \mathbf{k} \vert^\alpha$. That is not verified for surface waves on Earth, as a result of the competition of two restoring forces. Gravity dominates at large scales, whereas capillarity is the main restoring mechanism for wavelengths smaller typically than $15\,$mm for water. Experimentally, forcing low frequency gravity waves leads in the capillary wave range to power-law spectra of wave elevation whose exponents are in agreement to those predicted for pure capillary waves~\cite{Zakharov1967} both for the frequency spectrum $f^{-17/6}$~\cite{Falcon2007} and for the wavenumber spectrum  $k^{-15/4}$~\cite{Berhanu2013,Berhanu2018}, although the transition between gravity and capillary waves is not described by wave turbulence theory. Moreover, in such experiments, the degree of nonlinearity is not weak~\cite{Berhanu2018} and dissipation is significant~\cite{Deike2012,Deike2014b} even in the decade over which the predictions of weak wave turbulence are observed (typically between $20\,$Hz and $200$\,Hz). Recent direct numerical simulations~ \cite{Deike2014,Pan2014} have demonstrated  the relevance of weak wave turbulence theory for pure capillary waves in the limit of low dissipation and nonlinearity. Operating in microgravity conditions is required to study turbulence of pure capillary waves in laboratory experiments. Note that using two fluids of the same density leads to a different phenomenology~\cite{During2009}. A single study has been performed in parabolic flights to investigate turbulence of capillary waves inside a spherical container~\cite{FalconFalcon2009}. In this geometry surface waves propagate at the spherical air-water interface in absence of walls causing reflections or enhanced dissipation due to the contact line dynamics~\cite{Michel2016,Michel2017thesis}. Self-similar frequency spectra of wave elevation have been reported over two decades in satisfying agreement with the weak wave turbulence theory~\cite{FalconFalcon2009}. Nevertheless, due to the short duration ($22\,$s) of the zero gravity phases in parabolic flights, statistically steady states were not reached: in particular,  the behavior at low frequency was not accurately resolved and the transient motion of the fluid possibly interfered with the wave dynamics. {The observed power-law spectrum could correspond to a transient regime rather than to a capillary wave turbulence regime. To strengthen these previous observations}, we report in this letter experiments in the international space station (ISS), where the duration of the weightlessness is not limited. The low gravity condition is excellent with a root-mean-square (r.m.s.) acceleration due to external vibration less than $20 \times 10^{-6}\,g$ with $g=9.81\,$m$\cdot$s$^{-2}$, whereas the r.m.s. residual acceleration in parabolic flight is of order $5 \times 10^{-2}\,g$ is less. We use the FLUIDICS (Fluid Dynamics in Space) facility~\cite{Mignot2017} designed by {Airbus Defence and Space} and by the French spatial agency the CNES ("Centre National d'\'Etudes Spatiale") primarily to study the sloshing motion of liquid propellants of satellites~\cite{Dalmon2018,Dalmon2019}. A specific tank equipped with two fluid level sensors is dedicated to the study of capillary wave turbulence. The experiment was installed in the ISS and operated by ESA (European Space Agency) astronauts (Thomas Pesquet on May 3rd 2017, Paolo Nespoli on October 27th 2017,  Norishige Kanai on March 13th 2018 and Alexander Gerst on September 27th 2018). 

\begin{figure}
\onefigure[width=8.8cm]{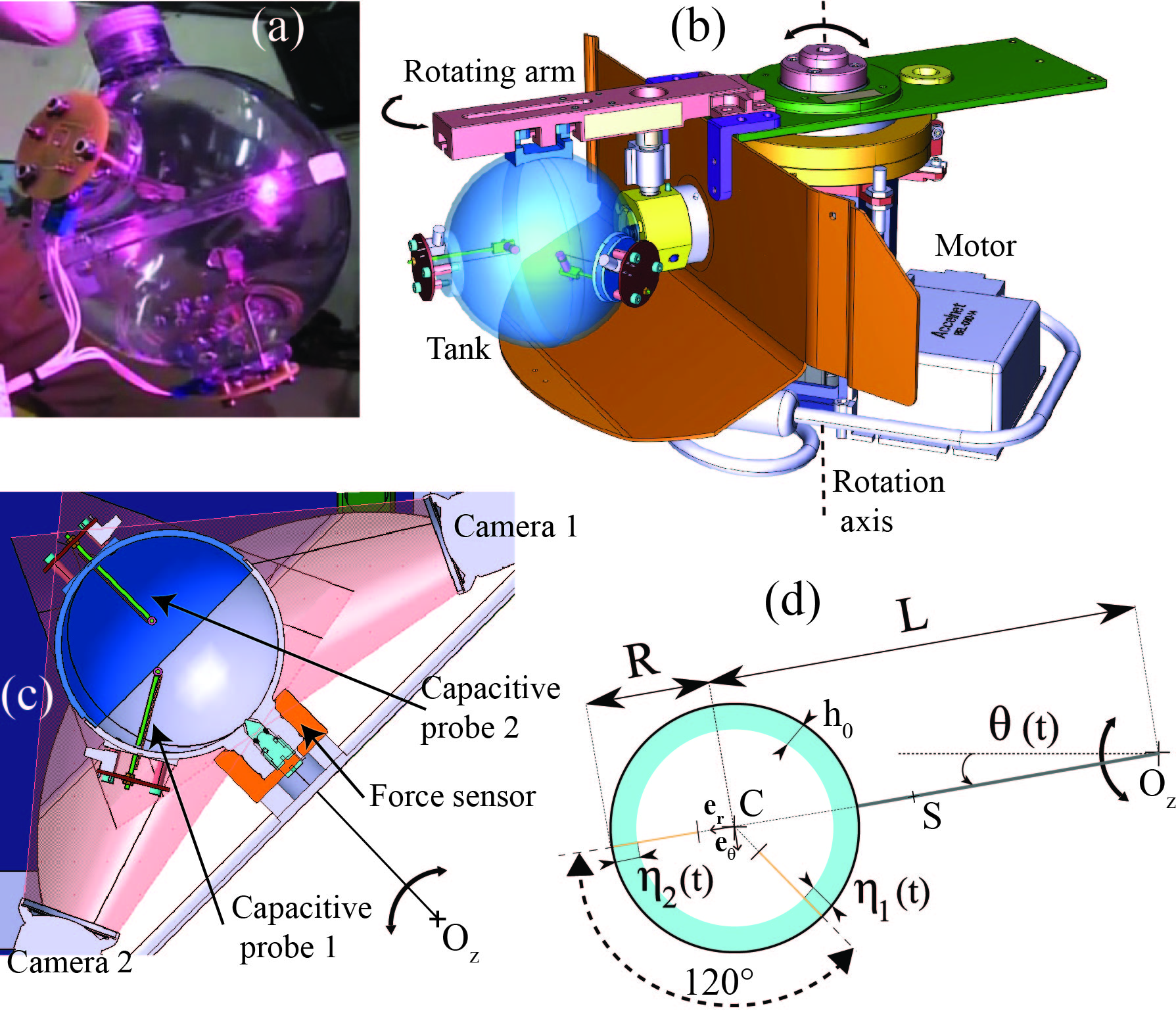}
\caption{(a) Picture of the experimental spherical tank. The cylinder visible on the top of the picture is the filling orifice and the fastening point. The two circular electronic boards measure the capacity for each fluid height probe. (b) Technical drawing of the experiment without, the acquisition system and the cameras ({Airbus Defence and Space}). (c) Horizontal cross section through the center of the tank. (d) Schematic view of the experiment (see text for definitions).}
\label{Fig1}
\end{figure}

\section{Experimental setup} The experimental setup is depicted in Fig.~\ref{Fig1}. A spherical container of internal radius $R=50\,$mm and made of polycarbonate is partially filled with $30$\,\% of water (mineral water  \textit{Luchon} \copyright) and with air at atmospheric pressure. In the ISS, the temperature is regulated around $22^\circ$\,C. The water density and the water/air surface tension are $\rho=998 $\,kg$\cdot$m$^{-3}$ and $\gamma=72$\,mN$\cdot$m$^{-1}$. The tank is rotated about the axis $O_z$ distant of  $L=175.4$\,mm with an oscillating angle $\theta(t)$ in order to generate waves at the air-water interface, $\theta(t)$ being either a sine function or a low-frequency noise. Two cameras image the transparent container in the rotating frame (see Fig.~\ref{Fig1} (c)). A force sensor ($S$ in Fig.~\ref{Fig1} (d)) located approximately at a distance of $74.8\,$mm from $C$ provides the three components of the force exerted by the tank on the rotating arm, so that the tank acceleration can be evaluated. Two capacitive probes~\cite{McGoldrick1971} (isolated copper wires of diameter 0.32 mm) measure the thickness of the  liquid layer, at two different positions $\eta_1 (t)$ and $\eta_2(t)$. More details on the experimental setup and on the data interfacing can be found in Ref.~\cite{Mignot2017}.
In weightlessness, the water wets the internal surface of the tank at rest, forming a large air bubble in the center and a homogeneous water shell of thickness $h_0=5.60\,$mm. This observation previously reported~\cite{FalconFalcon2009} can be understood by an energetic argument. Let $V_0$ denote the volume of water, $R_h = R-h_0$ the inner radius of the uniform shell and $R_0 = (3V_0 / 4\pi)^{1/3}$ the radius of a water sphere not in contact with the solid surface, and $\vartheta=78^\circ$ the static contact angle between water and polycarbonate~\cite{Cho2005}. The capillary potential energy of the wetting liquid shell is smaller than the inner liquid sphere, if $(R_h^2-R_0^2)< \cos\vartheta\,R^2$, \textit{i.e.} if the filling ratio $(R_0/R)^3 > 0.31$. Although the experimental value of this ratio $0.3$ is slightly smaller than the critical value, we observe at rest a homogeneously wetted sphere (see Fig.~\ref{Fig1} (a)). Nevertheless, the capillary forces are not sufficient to prevent some dewetting events when the tank is oscillated.\\
Considering now small oscillations of angular frequency $\omega$ of the wetting spherical shell the free-surface deformation $\eta$ can be decomposed using the spherical harmonics $Y_l^m$ and the amplitudes $C_{l,m}$:
$$\eta(t,\theta,\phi)=R_h+\sum_{l=0}^\infty \sum_{m=-l}^l C_{l,m}\,Y_l^m (\theta,\phi)\,\e^{i\,\omega t}.$$
With the assumption of potential flow, the dispersion relation can be derived~\cite{Lamb1932}:
\begin{eqnarray}\omega^2=\frac{\gamma}{\rho}\,\dfrac{l(l-1)(l+2)}{R_h^3}\,\dfrac{\left(\dfrac{R}{R_h}\right)^{2l+1}-1}{1+\dfrac{l}{l+1}\left(\dfrac{R}{R_h}\right)^{2l+1}}
\label{RDsphere}
\end{eqnarray}

When $l\geq 9$ (\textit{i.e.} $f=\omega/(2\pi)>3.7$\,Hz), the usual dispersion relation for capillary waves
\begin{eqnarray}
\omega^2=\frac{\gamma}{\rho} k^3 
\label{RD}
\end{eqnarray}
with $k = l/R_h$ is accurate with an error less than $5\,$\%. Note that this corresponds to a scale-invariant dispersion relation as assumed by the theory of weak wave turbulence.\\

\section{Forcing characterization} 
\begin{figure}
\onefigure[width=8.8cm]{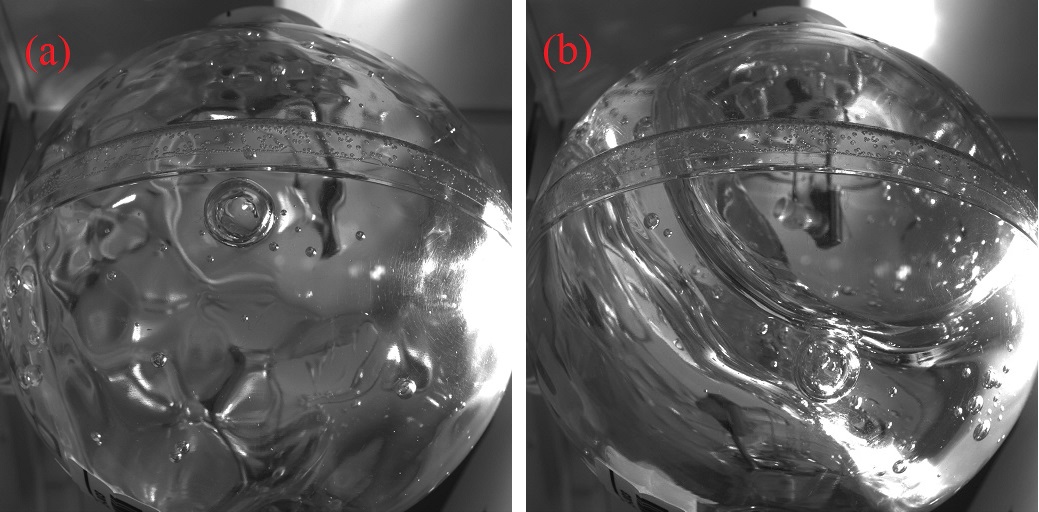}
\caption{Snapshots of the tank from camera 1, for a sinusoidal forcing with $F=1.75\,$Hz and $\theta_0=0.04$. Presence of air bubbles is noted. In the right image (b), a strong transient dewetting is visible on the top right corner.}
\label{Fig2}
\end{figure}

We consider first a sinusoidal forcing,  $\theta (t)=\theta_0\,\sin (2\pi\,F\,t)$, applied in most cases during $450\,$s. In the reference frame of the tank, the tangential and centrifugal accelerations at the center of the sphere read respectively $\mathbf{a_\theta}=-L\,\theta_0\,(2\pi\,F)^2\,\sin(2\pi\,F\,t) \,\mathbf{e_\theta}$ and $\mathbf{a_r}=-L\,\theta_0^2\,(2\pi\,F)^2\,\cos^2(2\pi\,F\,t) \,\mathbf{e_r} $. Therefore, to decrease the relative influence of the centrifugal force, which acts as an effective gravity, we restrict our study to the case of small oscillations, \textit{i.e.} $\theta_0 \leq 0.04$ (less than $2.5^\circ$). For a typical forcing $F=1.75\,$Hz and $\theta_0=0.04$, the liquid is strongly agitated and remains most of the time in contact with the surface of the sphere. However, a significant amount of bubbles of few millimeters are generated by the motion as it can be seen on the images from camera 1 (see Fig.~\ref{Fig2}). Morever a few dewetting events occur especially close to the tank fastening.\\
The second set of measurements corresponds to a random forcing with $\theta (t)=\theta_0\,\sin (\phi(t))$, where $\phi(t)$ is a random noise, band-pass filtered in the frequency interval $\Delta F=[F_1,F_2]$ with typically $F_1=1\,$Hz and $F_2=2\,$Hz and $\theta_0=0.04$\,. The visual observations are very similar. 

\begin{figure}
\onefigure[width=7.5cm]{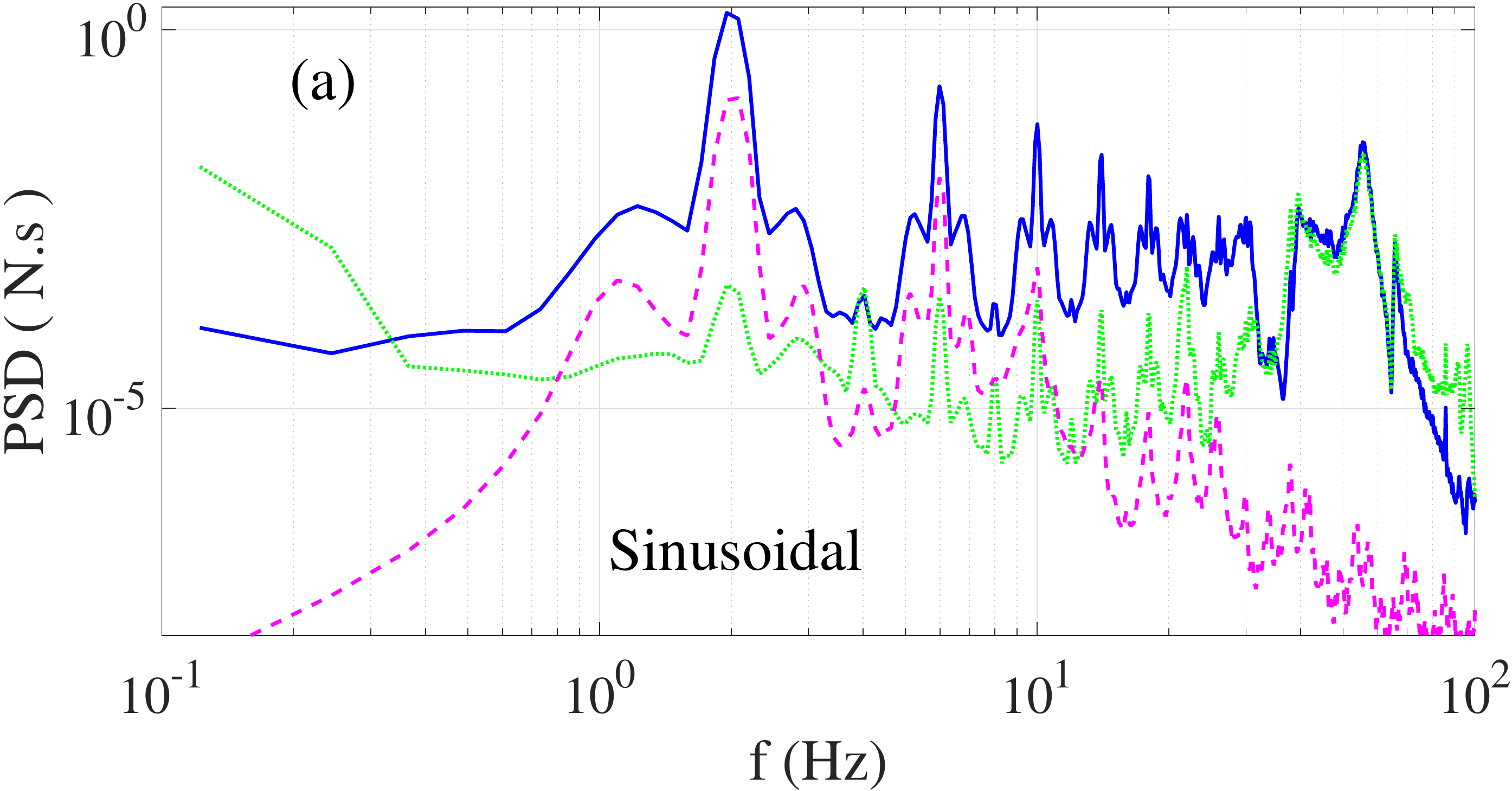}
\onefigure[width=7.5cm]{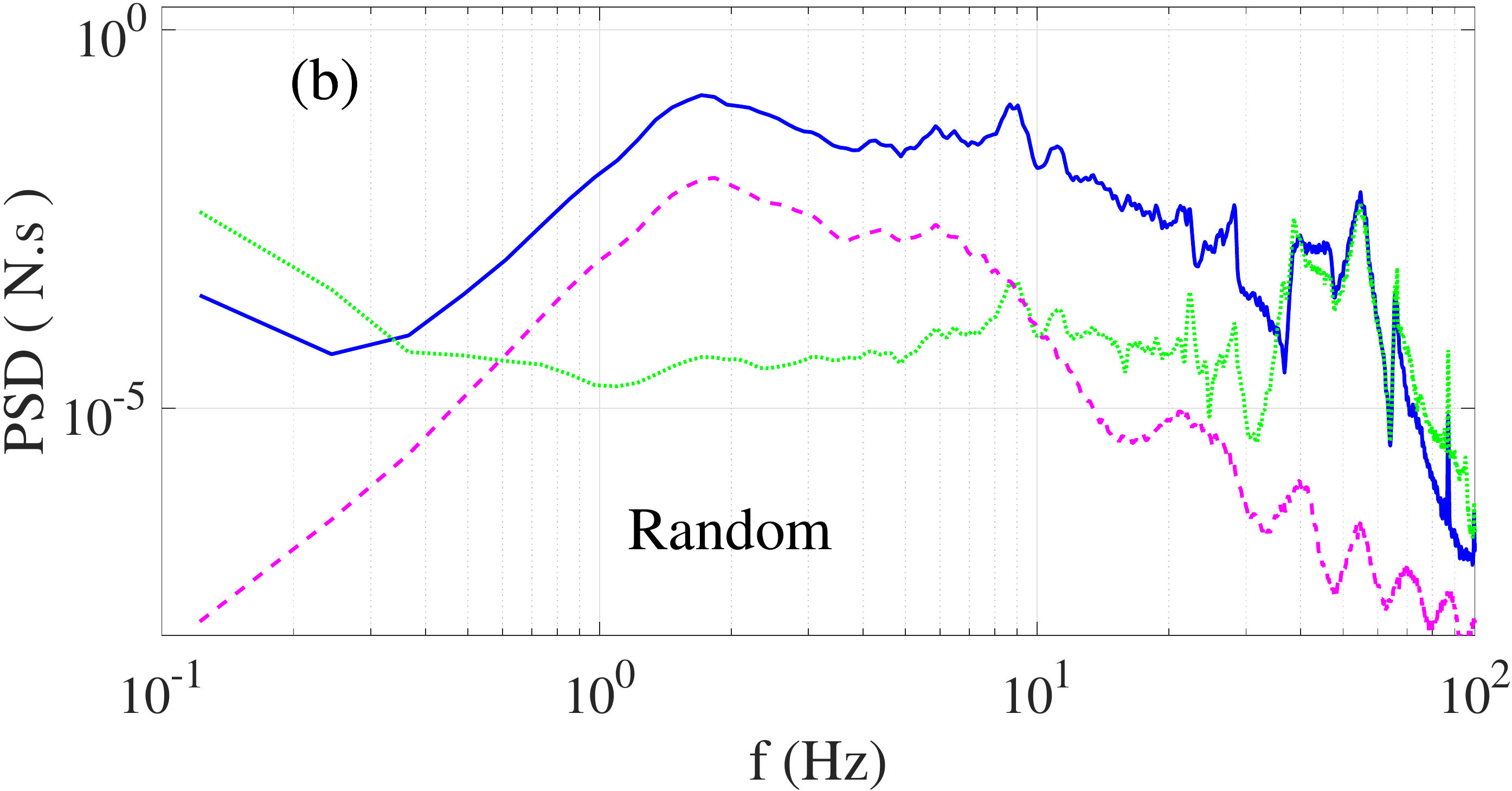}
\caption{Examples of Fourier forcing power spectra. Dashed magenta line, tangential force evaluated from $\ddot{\theta} (t)$.  Tangential force (continuous blue line) and radial force (dotted green line) are measured by the force sensor at $S$. (a) Sinusoidal forcing $\theta_0=0.04$ and $F=2\,$Hz. (b) Random forcing $\theta_0=0.04$ and $\Delta F=[0.5,2]$ \,Hz.}
\label{Fig3}
\end{figure}

The actual forcing can be characterized using the angular displacement of the motorized arm and the force sensor, power spectra being reported in Fig.~\ref{Fig3}. The arm follows satisfyingly the programmed instructions, although we notice significant  third and fifth harmonics for the sinusoidal forcing in the spectrum of the tangential force. This spectrum differs from the one from the force sensor, due to the inertia of the water during the motion and we observe a higher content between $10$ and $100$\,Hz. For the random forcing, the power spectrum is not flat in the expected range $[0.5,2]\,$Hz due to the mechanical response of the motor, with a maximum of excitation at about $1.7\,$Hz. The peak around 9 Hz in the force sensor data could be due to a sloshing eigenmode or to a vibration of the structure. For both sinusoidal and random forcing the spectra of the radial force is smaller by roughly a factor of $500$ for $f \in [1,30]\,$Hz. Moreover, the average value of the radial centrifugal acceleration at the furthest point of the tank from the rotation axis is $g_{eff}= 0.058\,$m$\cdot$s$^{-2}$ in the sinusoidal case with $\theta_0=0.04$ and $F=2\,$Hz, making this effective gravity negligible for frequencies $f\gtrsim 0.29$\,Hz. At the same location, the r.m.s. tangential acceleration is $a_{\theta}=1.4$\,m$\cdot$s$^{-2}$. In this setup in microgravity, deformation of the air-water interface can therefore be considered to be pure capillary waves, for frequencies larger than $1\,$Hz instead of about $20\,$ Hz on Earth. 


\section{Turbulent dynamics of the air-water interface} 
\begin{figure}
\onefigure[width=8.8cm]{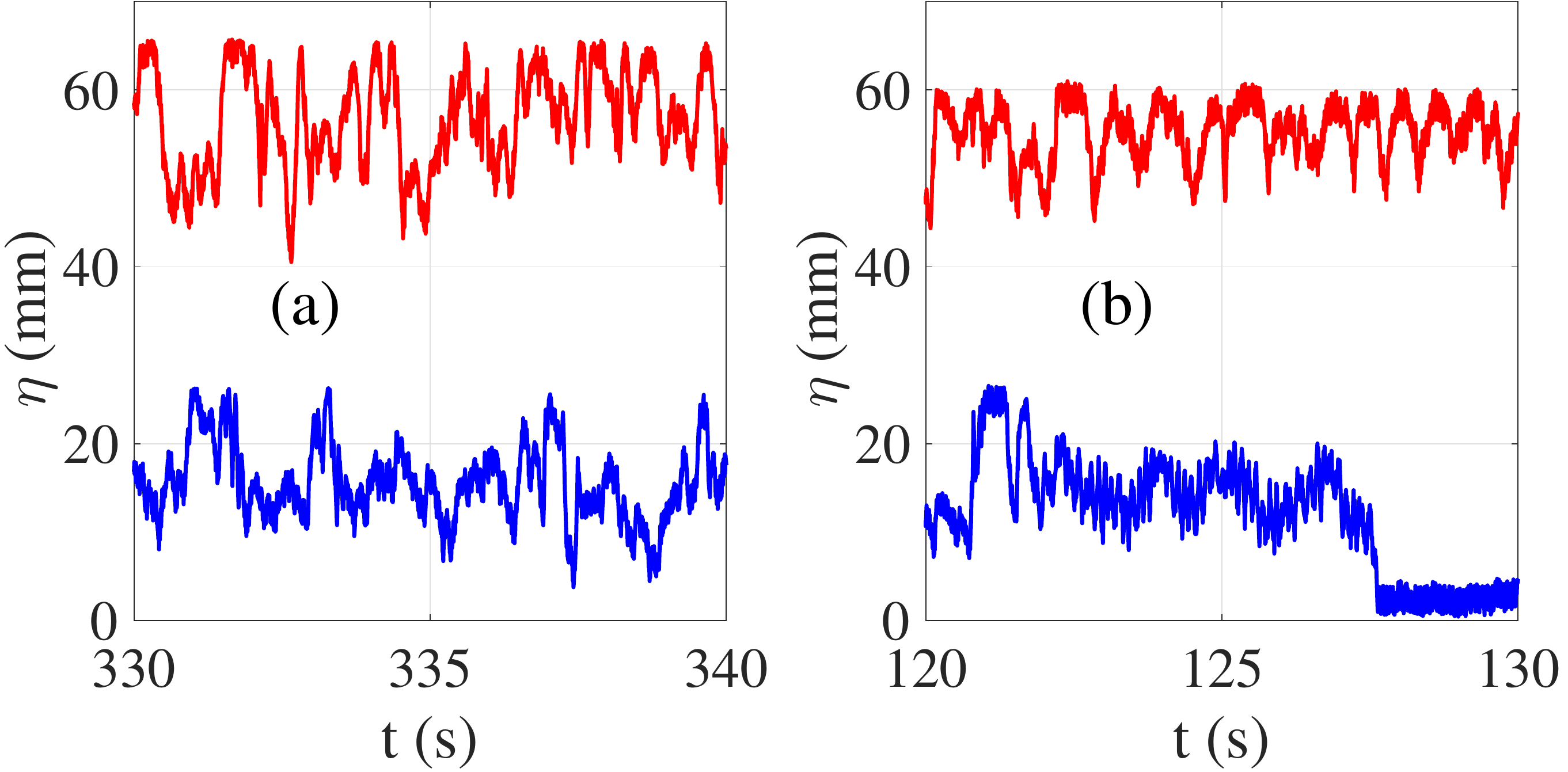}
\caption{Examples of temporal evolution of the local wave amplitude measured with the  capacitive probes. Blue (bottom), sensor 1, $\eta_1 (t)$. Red (top), sensor 2, $\eta_2 (t)$, the signal has been vertically shifted $35\,$mm for better visualization. (a) Sinusoidal forcing $\theta_0=0.04$ and $F=2\,$Hz. (b) Random forcing $\theta_0=0.04$ and $\Delta F=[0.5,2]$ \,Hz. At $t \gtrsim 127.5\,s$, a sharp decrease of $\eta_1 (t)$ is interpreted as a dewetting of the liquid from the tank wall at the location of the sensor 1. Note that sensor 2 is not affected.}
\label{Fig4}
\end{figure}

The fluctuations of the free-surface are monitored using the two capacitive probes. Examples of temporal evolutions are displayed in Fig.~\ref{Fig4}. The repetition of the sequences shows a good reproducibility. The sensor 1 $\eta_1(t)$ can be affected by the dewetting. As the position of the sensor 2 corresponds to the point of maximal centrifugal force, dewetting does not occur there and we favor this sensor in our further analysis. The sensor 2 $\eta_2(t)$ is sometimes saturated, which means that the sensor is completely immersed in water. However, the shape of the power spectra is not perturbed by these short saturations. For a sinusoidal forcing of frequency $F=2\,$Hz and amplitude $\theta_0=0.04$ corresponding to a forced wavenumber of $k \sim 130$ m$^{-1}$ we observe r.m.s. fluctuations of the free surface of order $\sigma_\eta \sim 5\,$mm. This order of magnitude close to the simple relation $a_t /(2\pi\,F)^2$ implies a relatively large value of the wave steepness $s \approx k \, \sigma_\eta \sim 0.6 $ using the linear dispersion relation Eq.~\ref{RD}. This high value of $s$ which quantifies the importance of nonlinear effects for surface waves shows that these experiments do not lie in the weakly nonlinear regime. However, the analytic solution of Crapper~\cite{Crapper1957} shows that the change of the dispersion relation of a plane monochromatic capillary wave does not exceed $10$\,\% at the maximal steepness value  $s=4.59$ corresponding to the trapping of air bubbles. Therefore, we use in the following the linear dispersion relation Eq.~\ref{RD}, which was experimentally verified in microgravity~\cite{FalconFalcon2009}. 

\begin{figure}
\onefigure[width=7.5cm]{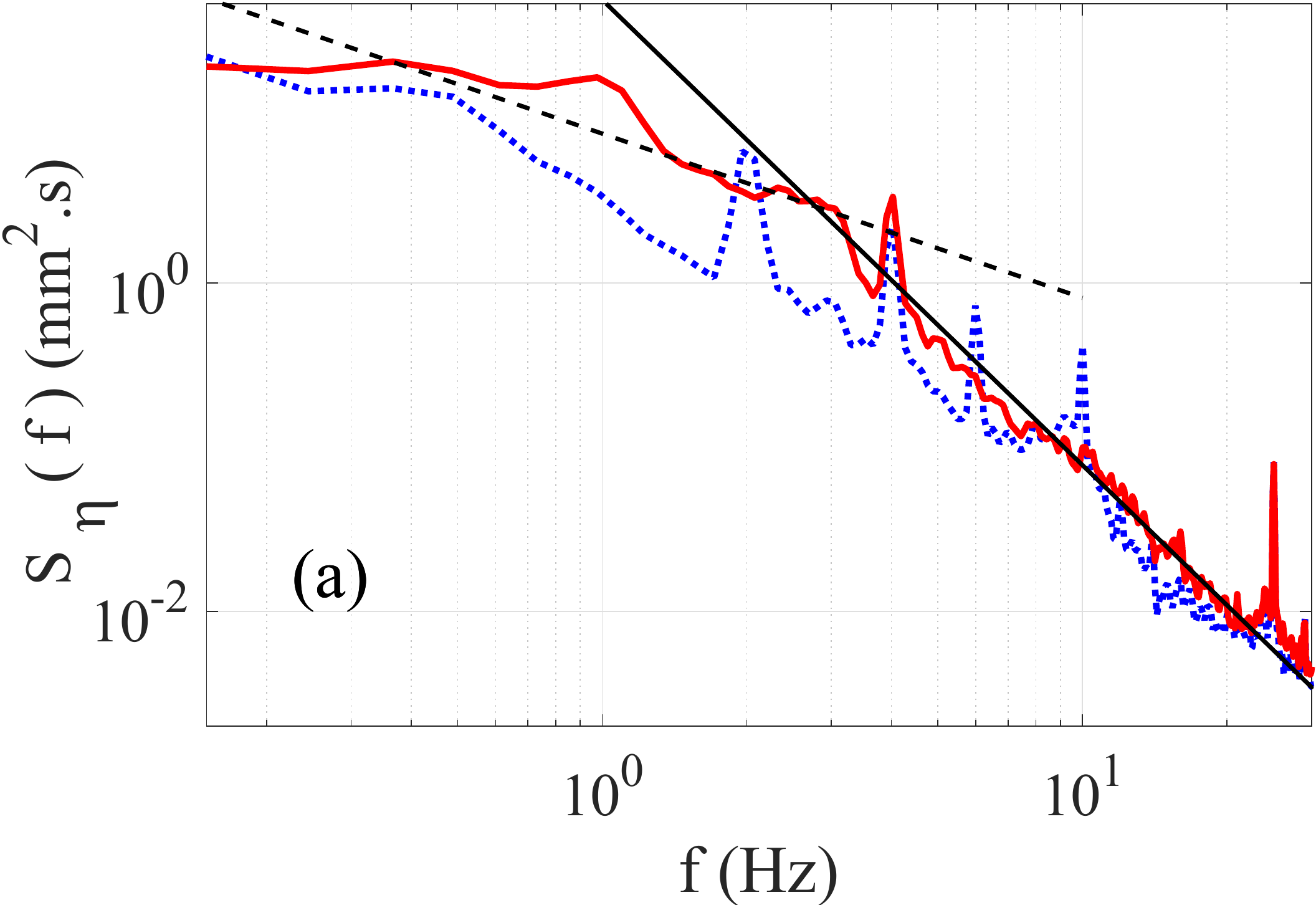}
\onefigure[width=7.5cm]{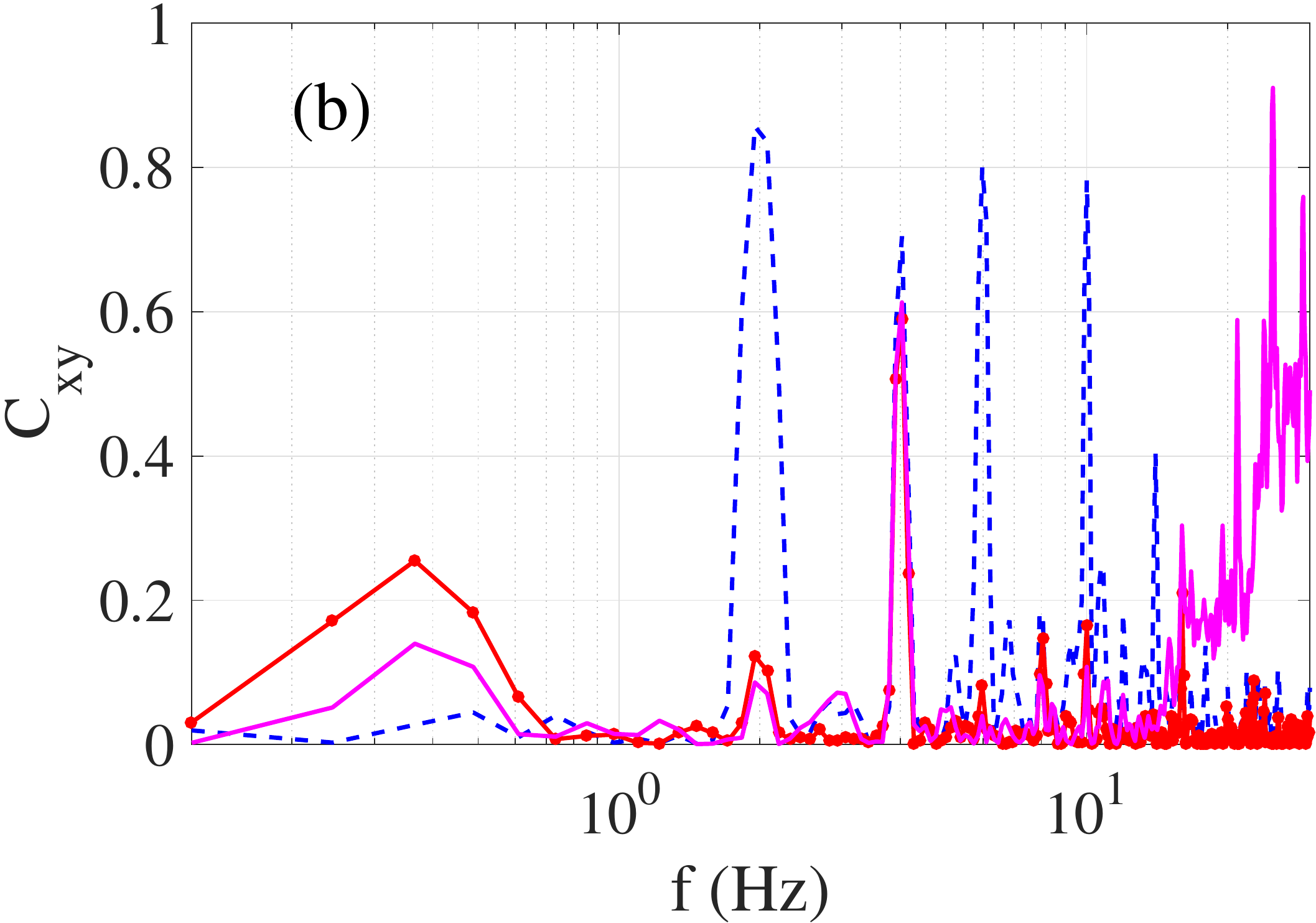}
\caption{(a) Power spectrum of wave elevation $S_{\eta_1} (f)$ (Blue dotted line) and  $S_{\eta_2} (f)$ (Red continuous line) for a sinusoidal forcing $\theta_0=0.04$ and $F=2\,$Hz.  The power-laws $f^{-17/6}$ and $f^{-1}$ are also indicated, respectively as a black continuous line and as a black dashed line (b) Magnitude-squared coherence functions $C_{xy}$ between $\eta_1$ and $F_t$ (blue dashed line), between $\eta_2$ and $F_t$ (red dotted continuous line) and between $\eta_1$ and $\eta_2$ (magenta continuous line).}  
\label{Fig5}
\end{figure}

The dynamics of the free-surface is characterized by the frequency power spectrum of wave elevation $S_\eta (f)$. The runs last for $450\,$s or more and the spectra are computed on a duration of $400\,$s (typically 20 times longer than for the parabolic flight experiments). Typical spectra for sensors $1$ and $2$ are plotted in Fig.~\ref{Fig5} (a) for a sinusoidal forcing of high amplitude ($F=2\,$Hz and $\theta_0=0.04 $). We observe for the sensor $2$, an absence of the forcing peak and a decay of the spectrum with frequency following fairly the power law predicted by the weak wave turbulence theory for capillary waves $S_\eta (f) \sim f^{-17/6}$~\cite{Zakharov1967}. Similarly to previous experiments operated in parabolic flights~\cite{FalconFalcon2009}, a capillary wave turbulent cascade is reported for a sinusoidal forcing of the tank, which is only observed for few experiments on Earth using a \textit{parametric} forcing in the capillary wave range ~\cite{Wright1996,Henry2000}. In contrast, when capillary waves are generated from nonlinear interactions of gravity waves, a random forcing appears necessary to generate the capillary wave power law spectrum~\cite{Falcon2007,Berhanu2013,Berhanu2018}. In the present experiment, we observe a peak at 4 Hz (the second harmonic of the forcing frequency) superposing on the turbulent cascade. To avoid a possible saturation, the sensitivity of the capacitive probes specially realized for this setup is lower than the ones used in our previous experiments on Earth or in the parabolic flights~\cite{FalconFalcon2009} and the spectrum reaches the electrical noise level typically for $f\geq 20 \,$ Hz. We display for information, the power-law $f^{-1}$ recently reported for turbulent capillary waves and interpreted as a statistical equilibrium of large scales~\cite{Michel2017}. In addition, we compute the magnitude-squared coherence function $C_{xy} (f)$, which indicates the correspondence degree between two temporal signals $x (t)$ and $y (t)$ at a given frequency $f$. $C_{xy}$ between $\eta_1 (t)$ and the tangential force $F_t$ measured with the force sensors in Fig.~\ref{Fig5} (b) shows that the local free-surface dynamics results mainly from the motion of the tank. In contrast, $C_{xy}$ between $\eta_2 (t)$ and $F_t$ is lower, except at the peak at $4$\,Hz, showing that the free-surface dynamics at the position of the sensor $2$ is weakly correlated to the forcing and results from the turbulent dynamics of the capillary waves. Similarly, the coherence between the two wave height signals is small.

\begin{figure}
\onefigure[width=7.5cm]{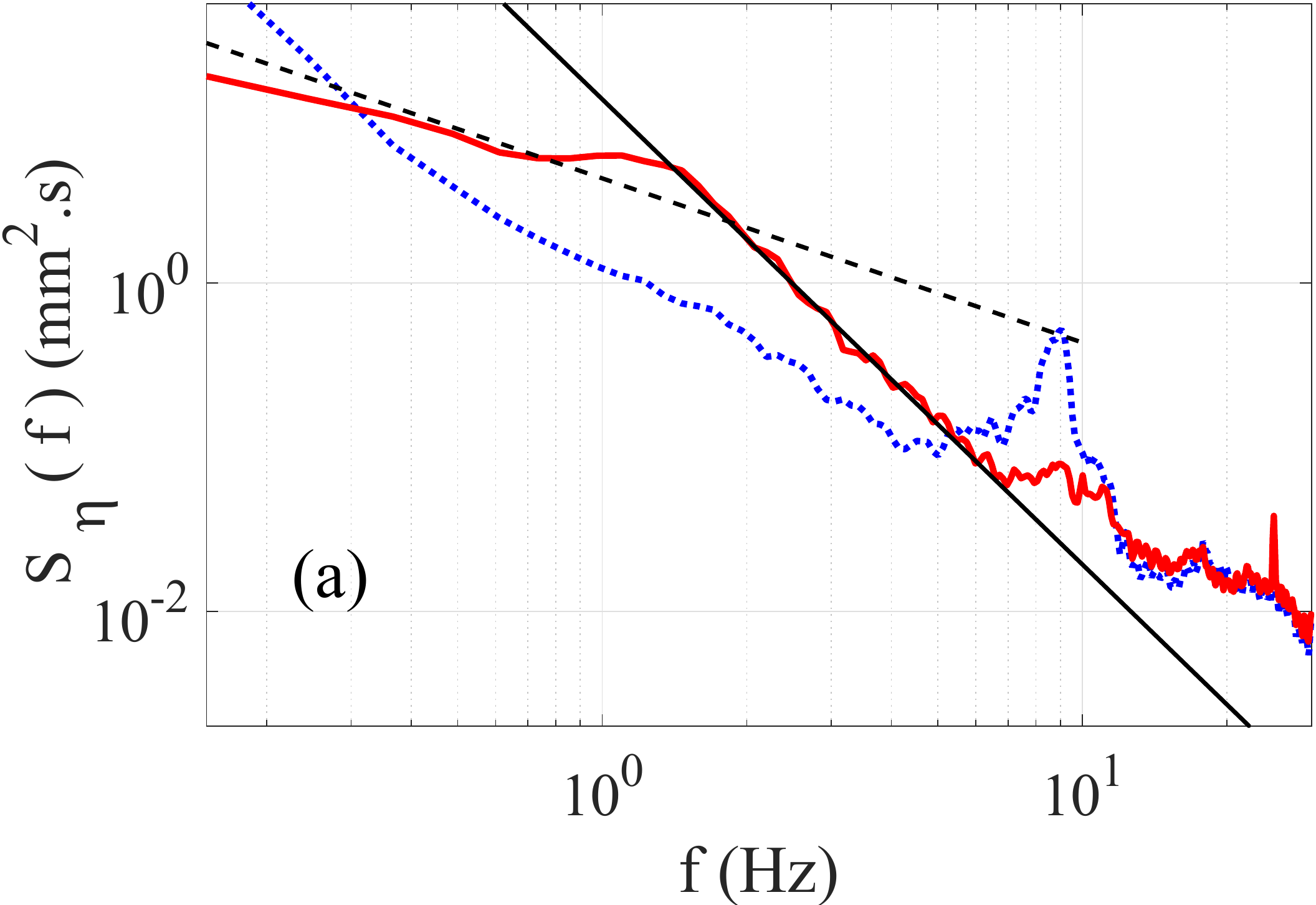}
\onefigure[width=7.5cm]{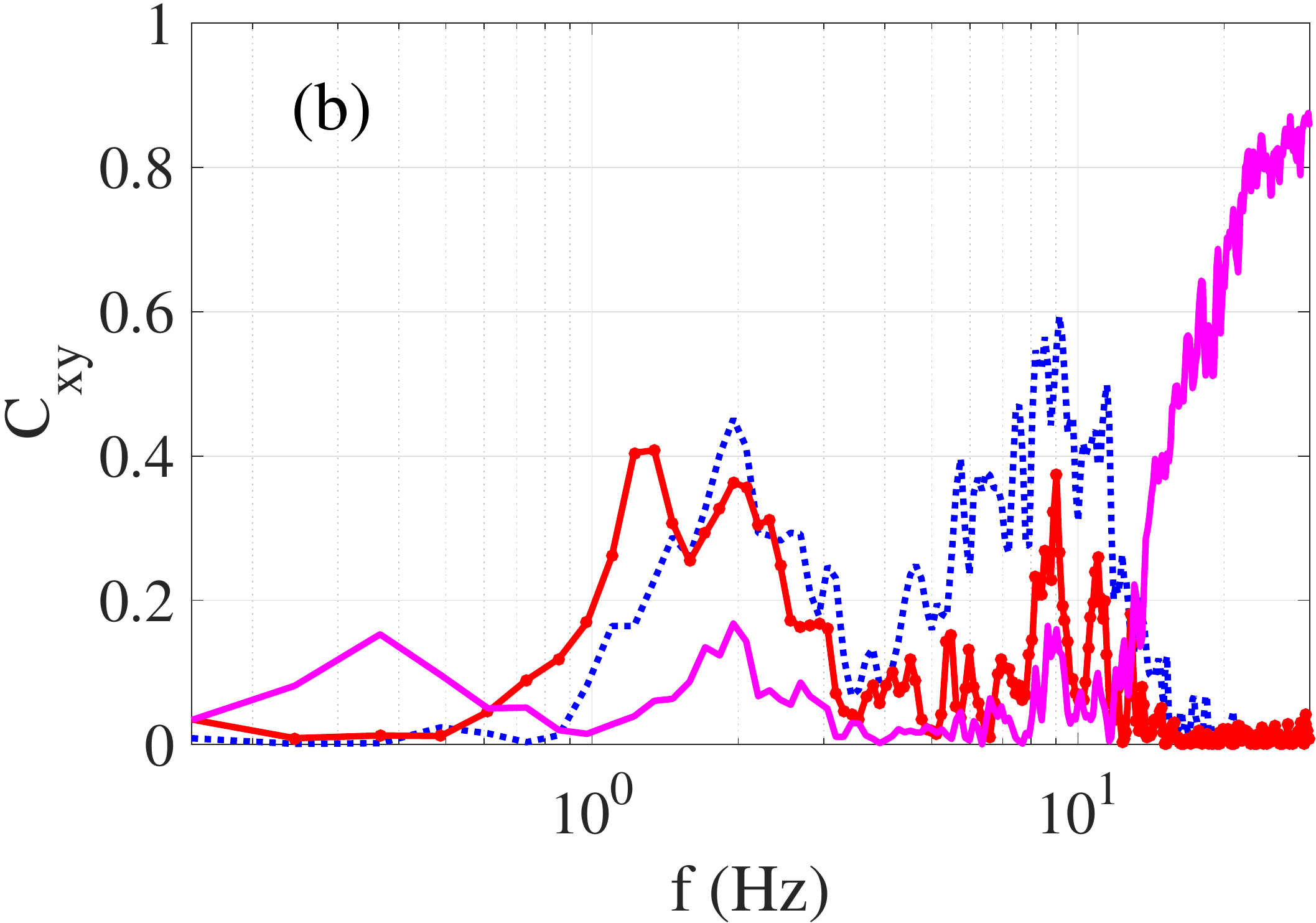}
\caption{(a) Power spectrum of wave elevation, $S_{\eta_1} (f)$ (Blue dotted line) and  $S_{\eta_2} (f)$ (Red continuous line) for a random forcing $\theta_0=0.04$, $F_c=1.25\,$Hz and $\Delta F=1.5$\,Hz. The power-laws $f^{-17/6}$ and $f^{-1}$ are also indicated, respectively as a black continuous line and as a black dashed line.  (b) Magnitude-squared coherence functions $C_{xy}$ between $\eta_1$ and $F_t$ (blue dashed line), between $\eta_2$ and $F_t$ (red dotted continuous line) and between $\eta_1$ and $\eta_2$ (magenta continuous line)}  
\label{Fig6}
\end{figure}

For runs operated with a random forcing, typically with an angular amplitude $\theta_0=0.04$ and a frequency range $f \in [0.5,2]$\,Hz, the spectrum displayed in Fig.~\ref{Fig6} (a) shows significant differences. The wave amplitude is lower at the maximal motor capacities, because the forcing is less coherent. The capillary wave turbulence power law is again observed for the second sensor without forcing peaks but on a narrower frequency range $f \in [1,8]\,$Hz. At frequencies lower than the forcing range, the spectrum is also compatible with the power law $f^{-1}$ expected for a statistical equilibrium of large scales~\cite{Michel2017}. At the location of the first sensor, we observe a significant peak at $9$\,Hz also visible on the data of the force sensor, which could be interpreted as a sloshing mode or a vibration of the structure. The coherence function in Fig.~\ref{Fig6} (b) between $\eta_1$ and $F_t$ displays indeed a quite high value around $9\,$Hz. The coherence between $\eta_2$ and $F_t$ is low outside frequencies belonging to the forcing range. The coherence between $\eta_1$ and $\eta_2$ remains also quite low, except for $f > 12\,$Hz, where the spectrum becomes dominated by the electrical noise, which is identical on both sensors.\\

\begin{figure}
\onefigure[width=8.cm]{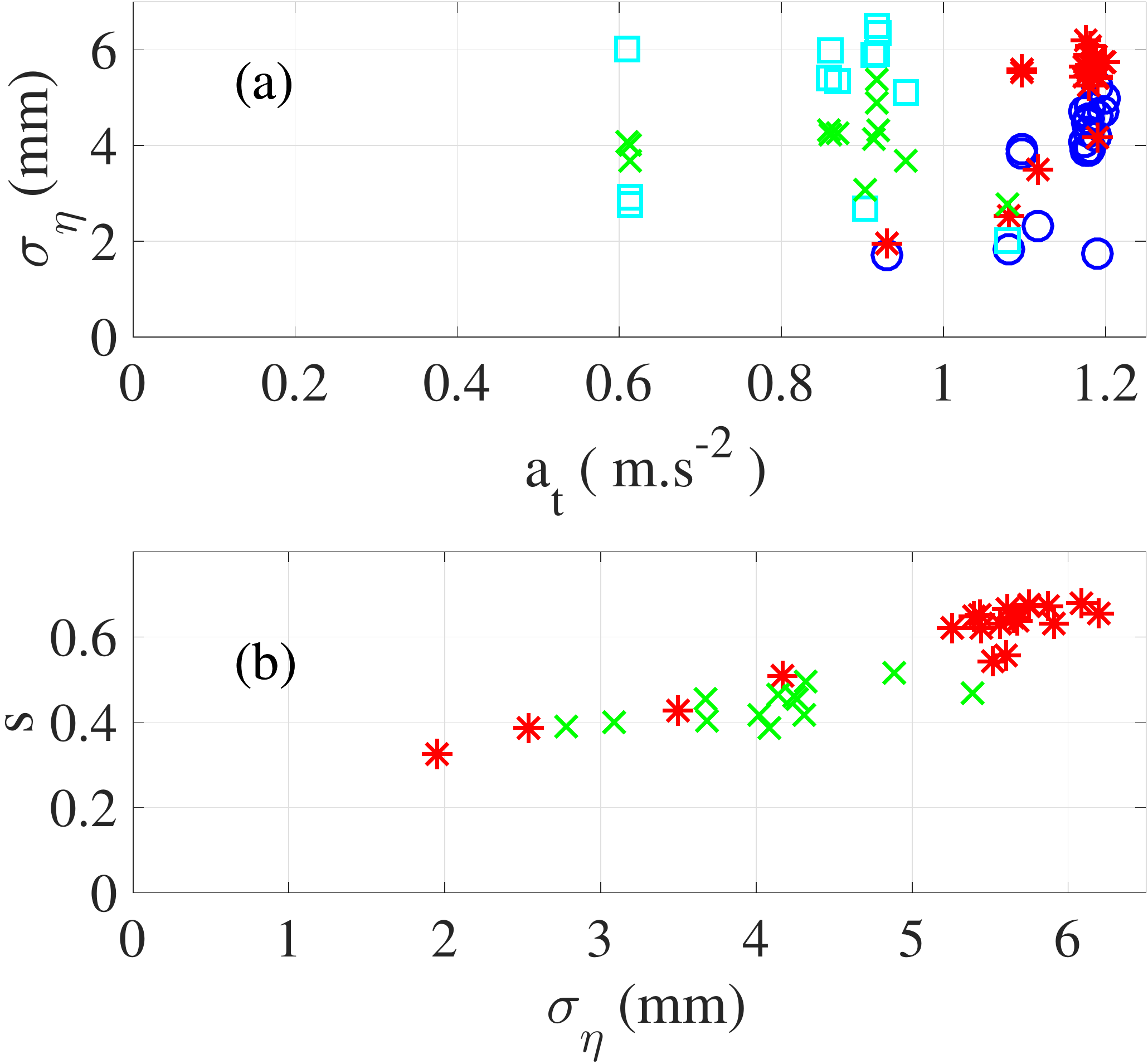}
\caption{(a) Standard deviation of the free surface as a function of the tangential forcing r.m.s acceleration. $\circ$ sensor $1$ sinusoidal forcing, $\ast$ sensor $2$ sinusoidal forcing, $\square$ sensor $1$ random forcing, $\times$ sensor $2$ random forcing. (b) For the sensor $2$ only, wave steepness $s \approx | \langle \nabla \eta_2 \rangle |$ as a function of the standard deviation of $\eta$. $s$ is derived from the spectrum $S_{\eta\,2} (f)$ and is a slowly growing function of $\sigma_\eta$.  $\ast$ random forcing,  $\times$ sensor $2$ random forcing.}
\label{Fig7}
\end{figure}

During the operations of the FLUIDICS experiment, 20 independent sinusoidal forcing runs ($F$ varying from $1.5$ to $6$\,Hz and $\theta_0$ varying from $0.01$ to $0.04$) and 13 random forcing runs  ($F_c$ varying from $1$ to $3.5$\,Hz, $\Delta F$ varying from $0.5$ to $1.75$\,Hz   and $\theta_0$ varying from $0.01$ to $0.04$) have been carried out, mostly for the highest possible torque of the motor, which is performing small oscillations. To summarize, the results from the different runs the standard deviation of wave elevation $\sigma_\eta$ is plotted in Fig.~\ref{Fig7} (a) as a function of the forcing tangential acceleration $a_t$ at the tank center location. For most of the runs, $a_t$ is of order $1.2\,$m$\cdot$s$^{-2}$ and $0.9\,$m$\cdot$s$^{-2}$  respectively for sinusoidal and random forcing, but the r.m.s. free-surface deformation remains close to $5\,$mm for both sensors. Using the linear dispersion relation Eq.~\ref{RD}, the spatial spectrum $S_\eta (k)$ is computed and the wave steepness from the signals of the second sensor can be evaluated as $s =\left(\int k^2 \, S_\eta (k) \, dk \right)^{1/2}=| \langle \nabla \eta_2 \rangle |$. Fig.~\ref{Fig7} (b) shows that $s$ is a slowly growing function of $\sigma_\eta$, with a typical wavenumber of {$k \sim s / \sigma_\eta \approx 110$\,m$^{-1}$ corresponding to a frequency of $1.6\,$Hz, which corresponds to the beginning of the capillary cascade. This estimation validates this indirect estimation of the wave steepness and shows that different runs of these experiments can be compared equivalently using the wave amplitude or the wave steepness.} Moreover, $s$ reaches quite large values, showing that the presented experiments are not performed in a weakly nonlinear regime.

\begin{figure}
\onefigure[width=8.8cm]{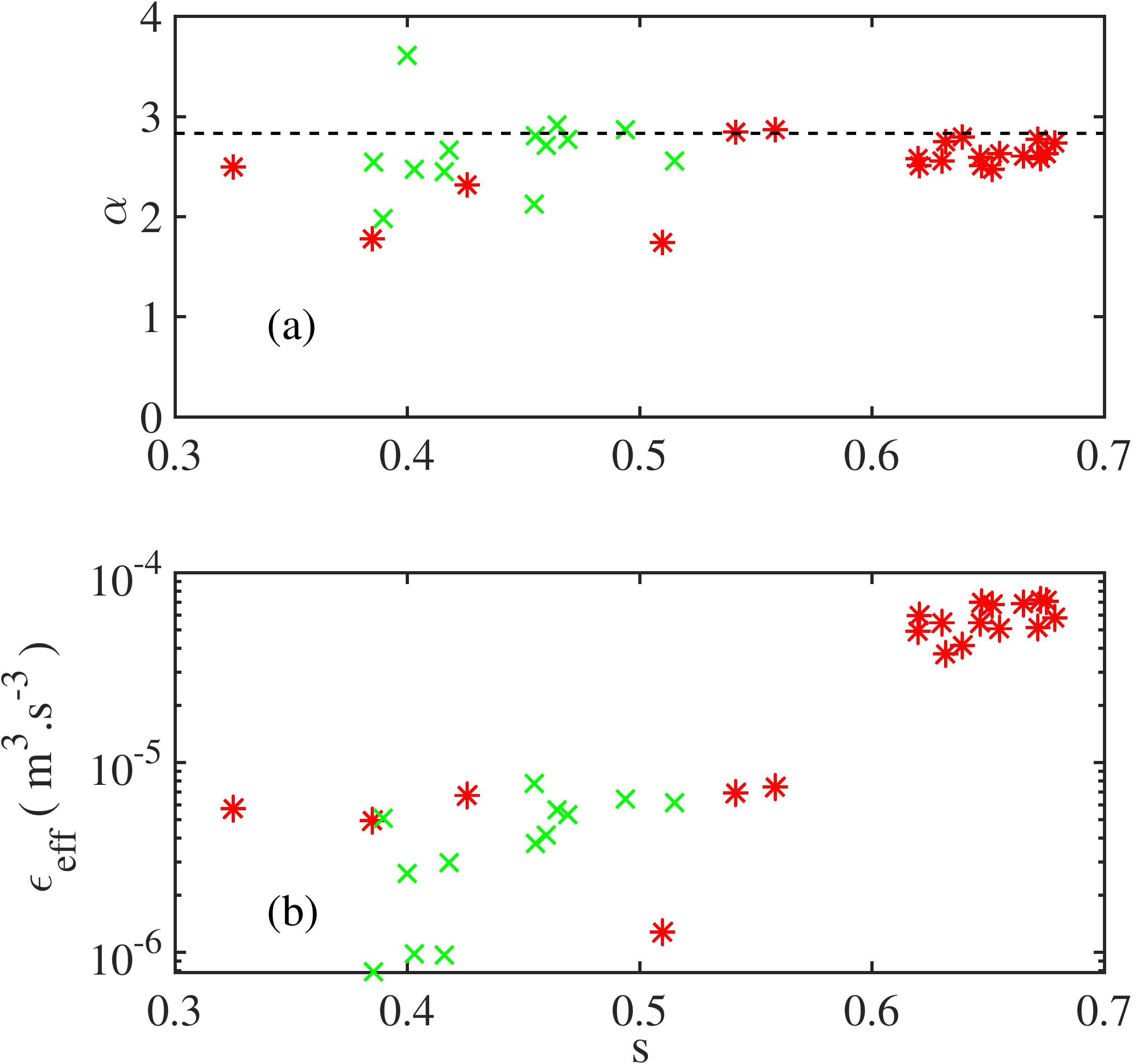}
\caption{For the sensor $2$ only, (a) fitted exponent of the power spectrum $S_{\eta} \sim f^{-\alpha}$ as a function of the steepness $s$. $\ast$ sinusoidal forcing,  $\times$ random forcing. The horizontal dashed line indicates the value predicted by the capillary wave turbulence theory $\alpha_{theo}=17/6$. (b) Effective energy flux $\epsilon_{eff}$ as a function of the steepness $s$.}
\label{Fig8}
\end{figure}

For all these runs, the power spectra are fitted in a variable frequency range as power laws $S_\eta = K_1 \,f^{-\alpha}$. The results are gathered in Fig.~\ref{Fig8} as a function of the wave steepness $s$. We observe first in panel (a), that for all measurements the exponent $\alpha$ is surprisingly robust, being close to $\alpha_{theo}$, especially at high $s$. The frequency interval of each fit is defined as the range of the capillary wave cascade \textit{i.e.} $f \in [2.5,20]$\,Hz for sinusoidal forcing and {$f \in [1.5,8]$\,Hz} for random forcing. To estimate the quality and the relevance of the fits, we compute a fit error ratio as $|S_\eta - K_1 \,f^{-\alpha}|/(K_1 \,f^{-\alpha})$ in the fit frequency range. For sinusoidal forcing, we report a fit error of order $15\,$\%. Note that at small values of $s$,  the fit error is significantly larger because it corresponds to runs operated at a smaller motor displacement, the corresponding spectra being not well described by a power law. For random forcing, we obtain a fit error of order $11\,$\%. The error is less than for sinusoidal forcing due to the absence of the peak at $4\,$Hz superposed on the cascade. 

Wave turbulence theory predicts an elevation power spectrum $S_\eta (\omega)=4 \pi/3\,C_{KZ}\,(\gamma/\rho)^{1/6}\,\epsilon^{1/2}\,\omega^{-17/6}$~\cite{Zakharov1967,Falcon2007,Berhanu2018}, where $\epsilon$ is the energy flux and the dimensionless constant $C_{KZ}$ can be analytically computed~\cite{Pushkarev2000,Pan2017} in the weakly nonlinear limit. To estimate the amplitude of the capillary wave turbulent cascade, our power spectra are fitted by $S_\eta (f)=K_2\, f^{-17/6}$. Although the present experiments do  not meet all the hypotheses of the theory (weak nonlinearity, negligible dissipation,…), an effective flux $\epsilon_{eff}=(K_2\,(2\,\pi)^{11/6}\,(\rho/\gamma)^{1/6})^2$ can be computed by assuming arbitrarily $4 \pi/3\,C_{KZ} =1$, thus providing a quantity of the same order of magnitude of the energy flux. Fig.~\ref{Fig8} (b) shows that this effective energy flux increases with the wave steepness. The values of $\epsilon_{eff}$ are comparable with the estimations from the dissipated power  by viscosity, which were obtained in a previous capillary wave turbulence experiment in a non weakly nonlinear regime~\cite{Berhanu2018}. From the scaling of the three-wave interaction collision integral, for a given value of the energy flux $\epsilon$ a critical wave-number can be defined $k_{NL} \approx \epsilon^{2/3}\,(\gamma/\rho)^{-1}$~\cite{Connaughton2003,Biven2001}. For $k<k_{NL}$ the linear time (the wave period) is smaller than the characteristic nonlinear time (the timescale of the wave interaction). Therefore, {capillary waves cannot propagate at too large scales, because they disappear due  to non-linear interactions on a time smaller than the period}. With the gravest experimental value of $\epsilon_{eff} \approx 7 \times 10^{-5}\,$m$^3$ $\cdot$s$^{-3}$, $k_{NL} \approx 23$\,m$^{-1}$. According to Eq.~\ref{RDsphere}, the largest possible mode $k_2=2/R_h \approx 45\,$m$^{-1}$ exceeds $k_{NL}$ and corresponds to a frequency of $f_2=0.22\,$Hz. Then, a breakdown of the {wave turbulence} due to a too large value of the energy flux does not occur in these experiments, {although the level of nonlinearity is significant}.

\section{Discussion} Using the FLUIDICS experiment, the turbulent fluctuations of an air-water interface have been investigated in weightlessness. Due to the excitation system by a rotating arm, the dynamics of the interface at the position of the first wave height sensor is perturbed by frequent dewetting events that affect the shape of the power spectra. However, this does not affect the signal at the location of the second wave height sensor that displays a surprisingly robust power law {in the frequency range $ [2.5,20]$\,Hz for sinusoidal forcing and $ [1.5,8]$\, Hz for random forcing. The exponents of these power-law spectra are in agreement with the prediction of capillary waves weak turbulence theory. This observation confirms the previous investigation of capillary wave turbulence in microgravity using parabolic flights~\cite{FalconFalcon2009}, which reported a power law spectrum of wave elevation in $f^{-3}$ close to the theoretical prediction in the frequency range $ [4,400]$\, Hz both for sinusoidal and random forcing. In these previous experiments, the wetting of ethanol on the glass cylindrical container insured indeed a better quality of the free-surface and the signal to noise ratio of the sensors was better. Nevertheless, the longer duration of the experiments in the ISS and the systematic investigation of forcing parameters insure that the power law spectrum is not caused by a transient behavior and can be attributed to capillary wave turbulence. Moreover, in the parabolic flight experiments~\cite{FalconFalcon2009}, the calibration of the capacitive probe was not provided and thus the amplitude of the waves in turbulent regimes was unknown. Here, we show that  the corresponding runs are obtained in fact} for relatively high values of the wave steepness $s \in [0.4,0.7]$, demonstrating that the capillary wave cascade can be observed outside the weakly nonlinear regime as stated in recent experiments~\cite{Berhanu2013,Berhanu2018}. {For a steepness level of order $0.1$, due to the nonlinear broadening of the dispersion relation, the effects of non resonant interactions are not negligible~\cite{Aubourg2016}. Moreover, the analytic derivation of the capillary wave turbulence spectra and of associated Kolmogorov-Zakharov constants~\cite{Pushkarev2000,Pan2017} may become invalid, as the theory requires a time scale separation between the nonlinear interaction time and the linear time. Here,} the value of the exponent of this strongly nonlinear wave turbulent cascade, close to $\alpha_{theo}=17/6$, might be explained by dimensional analysis arguments~\cite{Connaughton2003} in combination with the dispersion relation once we have supposed that the quadratic nonlinearity and thus three-wave processes dominate the wave dynamics~\cite{Berhanu2018}. Under strong agitation, the air-water interface in microgravity therefore presents a turbulent behavior, whose understanding remains incomplete and outside the validity domain of existing theories. Further experiments performed in microgravity may be useful to better characterize the strong capillary wave turbulence evidenced in this work.

\acknowledgments
We thank CNES (\textit{Centre National d'\'Etudes Spatiales}) for supporting this project. We acknowledge R. Pierre, S. Bonfanti, A. Papadopoulos and N. Bordes for the conception and the realization of the FLUIDICS experiment at Airbus Defence and Space. J. Mignot, L. Oro Marot, A. Llodra-Perez and M. Steckiewicz members of CADMOS at CNES have planned and supervised the operation of the runs. They have been also in charge of the communication with the ISS crew and the transmission of data. E. F. and S. F. thank the funding of ANR-17-CE30-0004 Dysturb.

\end{document}